\documentclass[11pt]{article}
\usepackage{jheppub}
\usepackage{amsmath,amssymb,amsfonts,graphicx}
\usepackage{amsthm}

\newcommand{\be}{\begin{equation}}
\newcommand{\ee}{\end{equation}}
\newcommand{\bea}{\begin{eqnarray}}
\newcommand{\eea}{\end{eqnarray}}
\newcommand{\beas}{\begin{eqnarray*}}
\newcommand{\eeas}{\end{eqnarray*}}
\newcommand{\ba}{\begin{array}}
\newcommand{\ea}{\end{array}}
\newcommand{\tr}{{\rm tr}}

\renewcommand*\d[2][]{%
	\mathrm{d}%
	\ifx\relax#1\relax\else
	\rule{-0.02em}{1.5ex}^{#1}\rule{0.08em}{0ex}\!
	\fi
	#2\,
}

\author[1]{Brian Swingle,}
\author[2]{Mark Van Raamsdonk}

 \affiliation[1]{Department of Physics, Brandeis University, Waltham, MA 02453, USA}
\affiliation[2]{Department of Physics and Astronomy, University of British Columbia, Vancouver, B.C.\ V6T 1Z1, Canada.}

\emailAdd{bswingle@brandeis.edu}
\emailAdd{mav@phas.ubc.ca}

\begin{document}

\title{Enhanced Negative Energy with a Massless Dirac Field}

\abstract{Motivated by traversable wormhole constructions that require large amounts of negative energy, we explore constraints on the amount of negative energy that can be carried by a free Dirac field in a slab-shaped region between two parallel spatial planes. Specifically, we ask what is the minimum possible uniform energy density that can exist at some time, considering all possible states and all possibilities for the physics outside the slab. 
The vacuum state where we identify the two sides of the slab with antiperiodic boundary conditions gives one possible state with uniform negative energy, but we argue that states with more negative energy exist above 1+1 dimensions. Technically, we reduce the problem to studying a massive Dirac field on an interval in 1+1 dimensions and numerically search for states with uniform energy density in a lattice regulated model. We succeed in finding states with enhanced negative energy (relative to the antiperiodic vacuum) which also appear to have a sensible continuum limit. Our results for the mass-dependence of the minimum uniform energy density in 1+1 dimensions  suggest that for a 3+1 dimensional massless Dirac fermion, it is possible to have states with arbitrarily large uniform negative energy density in an arbitrarily wide slab.}

\maketitle

\section{Introduction}

For any state of a Poincar\'e invariant quantum field theory on a patch of Minkowski space, there is some stress-energy tensor expectation value $t_{AB} = \langle T_{AB} (x) \rangle$ which can be defined relative to the Minkowski space vacuum. However, not every $t_{AB}$ can arise from a valid quantum field theory state. The stress-energy tensor obeys various constraints, for example the local conservation relation $\partial_A t^{A} {}_B = 0$ or tracelessness $t_A^A = 0$ in the case of a conformal field theory. An interesting set of nonlocal constraints restricts the amount of negative energy that is allowed. The Averaged Null Energy Condition (ANEC) states that for any valid state of the field theory on Minkowski space, the integral $\int d \lambda T_{AB} k^A k^B$ over a complete null geodesic must be positive, where $k$ is the tangent vector along the geodesic parameterized by $\lambda$. This has been proven recently using quantum information methods~\cite{faulkner_anec_2016}. Various other energy inequalities include the Quantum Energy Inequalities (see \cite{Fewster:2012yh} for a review) and the Quantum Null Energy Condition~\cite{bousso_qfocusing_2015,bousso_qnec_2015,wall_lowerbound_2017,balakrishnan_qnec_2019}.

In situations without complete null geodesics (e.g. when we have boundaries/interfaces or the quantum field theory state is defined only on a patch of Minkowski spacetime), it is less clear whether similar constraints on the amount of negative energy exist. In this paper, we investigate the amount of negative energy that can exist within a slab of width $L$ for a conformal field theory in $d$ spacetime dimensions, considering all possible extensions of the physics of the field theory outside the slab. This includes the case of the quantum field theory on Minkowski space, theories with various boundary conditions at the edges of the slab, or theories where the field theory degrees of freedom on the slab are coupled to different quantum field theories outside via some interface. Essentially, we are exploring the possible uniform energy densities that can arise when allowing all possible states and boundary conditions for the fields.

In \cite{may_large_neg_en_2021}, this question was considered in the context of holographic conformal field theories. There, for $d=3$ or $d=4$ (higher $d$ were not considered), it was found that the integrated null energy between edges of the slab, or the energy density for states with uniform energy density on the slab, is unbounded from below. Since the holographic model considered in \cite{may_large_neg_en_2021} is a ``bottom up'' model, based on a simple bulk gravitational theory without a specified UV completion, it does not precisely define a specific quantum field theory. In this paper, we ask whether similar results can be obtained in a simple conformal field theory that can be studied directly, the free Dirac fermion.

We consider the Dirac fermion in $d$ spacetime dimensions on a slab geometry where one of the spatial directions is an open interval of fixed length $L$ and the remaining spatial directions are translation-invariant (either periodic or infinite). In this case, the degrees of freedom may be divided into decoupled sectors labeled by transverse momentum $\vec{p}$. Each of these sectors is equivalent to a 1+1 dimensional Dirac fermion with mass $|\vec{p}|$. 

We introduce a simple and direct method to determine which energy distributions are possible for such 1+1 dimension fermions. First, we regulate the theory by defining it as the continuum limit of a lattice model with finite dimensional Hilbert space and a Hamiltonian that is a sum of terms $H_a$ associated with various sites. In the lattice theory, if a state with a certain distribution of energy exists, then there must also be a state with this distribution of energy that maximizes the von Neumann entropy $-\tr(\rho \log \rho)$. Thus, when searching for states with a given energy distribution, it is sufficient to restrict to such entropy-maximizing states. As we review below, these states take a particularly simple form, 
\be
\log \rho = - \sum_a \beta_a H_a + \text{constant},
\ee
where the modular Hamiltonian (the logarithm of the density matrix, $K = - \log \rho$) is local, i.e. it is a linear combination of the operators giving the energy associated with various sites in the lattice. In the continuum limit, we can describe these states as having a local modular Hamiltonian
\be
K = \int dx \beta(x) T_{00}(x) \; .
\ee

It is straightforward to numerically compute the distribution of site energies $E_a = \langle H_a \rangle$ for any parameters ${\beta_a}$. Using numerical optimization techniques, we can also start with a desired distribution of energies and search for a configuration of $\beta$s that achieves this. With this numerical approach, we investigate the minimum possible uniform energy density for 1+1 dimensional Dirac fermions of mass $m$ on an interval of width $L$ and use these results to study the possible energy densities for higher-dimensional massless Dirac fermions.

In 1+1 dimensions, our numerical results suggest that the most negative possible uniform energy density for massless Dirac fermions on an interval of width $L$ is the value obtained when the two ends of the interval are identified, with antiperiodic boundary conditions for the fermions. We are also able to prove this directly using the Quantum Energy Inequalities (appendix D). However, in the massive case, we find states with energy densities that are lower than the result for antiperiodic boundary conditions. These results are displayed in Figure \ref{fig:dataAP}. While the Casimir energy with antiperiodic boudary conditions falls off exponentially in the mass, the numerical results suggest that the minimum possible uniform energy density falls off only as a power of the mass.

Placing the various transverse momentum modes of higher-dimensional massless Dirac fermions into these energy-minimizing states (for $m = |p|$), we find states with negative energy densities that are enhanced compared to the Casimir energy density with antiperiodic boundary conditions.  

In 2+1 dimensions, we find states for the massless Dirac fermion on a slab of width $L$ for which the negative energy density exceeds that of the vacuum with antiperiodic boundary conditions by some finite factor. 

In 3+1 dimensions, for the theory on a slab of fixed width $L$, our numerical results suggest that there exists states with uniform negative energy densities of arbitrarily large magnitude (relative to the Minkowski space vacuum). In other words, after fixing the width of the slab to some arbitrary value and exploring states with a uniform energy density, there appears to be no lower bound on the energy density, even after subtracting off the Minkowski space energy density. 

The enhanced negative energy states we find (for $m >0$ in 1+1 dimensions and in higher dimensions) have a stress tensor component $T_{zz}$ that depends on $z$; by the conservation relation, this implies a time-dependent stress-energy tensor in which there is a flow of negative energy out towards the boundaries of the slab after the initial time.

\subsubsection*{Application to wormholes and cosmology}

For quantum field theories coupled to gravity, the existence or not of various types of negative energy has implications for whether or not certain types of solutions to Einstein's equations are physical. For example, certain traversable wormhole geometries require the existence of negative integrated null energy between the two sides of the wormhole. 

In \cite{vr_cosmo_confinement_2021,antonini_cosmo_vac_2022}, we argued that certain microscopic models of big bang cosmology could be defined in string theory via a specific class of holographic field theories. The viability of this construction seems to require the existence of asymptotically AdS planar traversable wormhole geometries (related to the cosmological spacetimes by double analytic continuation). To support such solutions, large amounts of negative  energy are required. Since the wormhole geometries are conformal to a 3+1 dimensional slab, the results of this paper suggest that rather ordinary quantum field theories might have states with the appropriate energy densities to support these wormhole solutions. Note, however, that the states required to support these wormhole solutions should also have 2+1 dimensional Poicar\'e invariance. The states we find do not have this property (we did not impose it as a constraint), so further investigation is required to understand whether states with unbounded negative energies and this Poincar\'e invariance exist as they do in the holographic models of \cite{may_large_neg_en_2021}.

\subsubsection*{Outline}

Here is the plan for the remainder of the paper. In section 2, we review the reduction of a (d-1)+1 dimensional Dirac fermion to a collection of 1+1 dimensional Dirac fermions with various masses. We describe the lattice regularization that allows us to define these 1+1 dimensional theories via a sequence of theories with finite-dimensional Hilbert space. In section 3, we describe our numerical procedure to determine whether a certain spatial distribution of energies is allowed in a given lattice model, by exploring the space of entropy-maximizing states with local modular Hamiltonians. In section 4, we describe our results for the allowed energy densities in various dimensions. We end in section 5 with a discussion.

\section{Basic setup}
\label{sec:setup}

%{\bf Below is a possible update for this section.}

Consider a conformal field theory in $d$ spacetime dimensions on a slab of width $L$ in one of the spatial dimensions (i.e. a subset $z \in (0,L)$ of Minkowski space) with unspecified boundary physics preserving $(d-1)$-dimensional Poincar\'e invariance. This could involve certain boundary conditions for the fields at the edges of the slab, or a coupling to additional degrees of freedom outside the slab. We denote the coordinate between the boundaries of the slab by $x^{d-1} = z$; the transverse coordinates correspond $\mu,\nu=0,\cdots,d-2$. 

We would like to consider all possible states of the theory with uniform energy density in the region $z \in (0,L)$ and ask what is the minimum possible value of this energy density, relative to the energy density corresponding to the Minkowski space vacuum.

\subsection{The Dirac Fermion}

For our investigation, we consider free massless Dirac fermions in one, two, and three spatial dimensions. We consider these theories on an open interval in one of the spatial directions, considering the theory to be a subsystem of a larger physical system. For the other spatial directions, we assume either periodic or infinitely extended directions with translation invariance.

The continuum model is defined by the usual Dirac action with Lagrangian density
\be
{\cal L} = i \bar{\psi} \Gamma^A \partial_A \psi \; ,
\ee
with two-component spinors in 1+1 and 2+1 dimensions and four-component spinors in 3+1 dimensions.

Letting $\vec{y}$ and $\vec{p}$ denote spatial position and momentum transverse to the interval, and writing
\be
\psi = \psi_{\vec{p}} \, e^{i p \cdot y},
\ee
we find that each momentum mode $\psi_{\vec{p}}$ is an independent physical system equivalent to a Dirac fermion in 1+1 dimensions with mass $|\vec{p}|$
\be
{\cal L}_{\vec{p}} = i \bar{\psi}_{\vec{p}} (\gamma^0 \partial_0 + \gamma^z \partial_z) \psi_{\vec{p}} + |\vec{p}| \bar{\psi}_{\vec{p}} \psi_{\vec{p}} \; ,
\ee
or in the 3+1 dimensional case to two independent 1+1 dimensional Dirac fermions with mass $|\vec{p}|$. This reduction is reviewed in detail in Appendix \ref{sec:fermions}. 

\subsection{The lattice model}

Given a Dirac fermion in $1+1$ dimensions on an open interval of length $L$, we would like to understand what is the minimum possible uniform energy density for any possible way of completing the physical system outside the interval. More generally, we can ask which distributions of energy $\langle T_{00}(x) \rangle$ are possible.

Since we are considering a quantum field theory on an open region, general states are most appropriately described as maps from the algebra of bounded operators on the region to $\mathbb{C}$. To be more concrete, we will make use of a lattice regularization of the field theory and consider quantum field theory states to be defined via sequences of density matrices on successively finer lattices.

\subsubsection*{Lattice fermions}

For our lattice regularization, we consider a model with two fermions on each site (related to the two components of the Dirac fermion), with Hamiltonian of the form
\be
H = A_i c^\dagger_n \sigma^i c_n + {{\cal R}_i \over 2} (c^\dagger_{n+1} \sigma^i c_n + c^\dagger_{n} \sigma^i c_{n+1}) + {{\cal I}_i \over 2} (i c^\dagger_{n+1} \sigma^i c_n - i c^\dagger_{n} \sigma^i c_{n+1}),
\ee
where $\sigma^i$ are Pauli matrices, $c_n^\dagger = (a_n^\dagger, b_n^\dagger)$ represent the fermion creation operators on each site, and $A_i$, ${\cal R}_i$, and ${\cal I}_i$ are real constants. Under $SU(2)$ rotations mixing the components of $c$, the parameters of the Hamiltonian transform as three-dimensional vectors, so the physical properties of the model depend only on rotational invariants made from the three vectors.

For an infinite or periodic one-dimensional lattice, we can diagonalize the Hamiltonian via states built from creation operators
\be
c^\dagger_{k,\alpha} = \sum_n e^{i k n} B_{\alpha \beta} c^\dagger_{n, \beta} \; .
\ee
for which
\be
[H, c^\dagger_{k,\alpha}] = E_\alpha(k) c^\dagger_{k,\alpha} \; .
\ee
We can choose the distinct momenta to lie in the range $k \in (-\pi,\pi]$; in the periodic case, we have the further restriction to discrete momenta $k = 2 \pi n / N$, while to model fermions with antiperiodic boundary conditions, we take $k = 2 \pi (n+1/2)/N$.

We find two modes for each $k$ that we can label as $\alpha = \pm$, with dispersion relation\footnote{Including additional terms $A_0 c^\dagger_n c_n + {1 \over 2} ({\cal R}_0 + i {\cal I}_0) c^\dagger_{n+1} c_n + {1 \over 2} ({\cal R}_0 - i {\cal I}_0) c^\dagger_{n} c_{n+1}$ in the Hamiltonian would add a term $A_0 + {\cal R}_0 \cos k + {\cal I}_0 \sin k$ to the dispersion relation, but we exclude this when targeting a Lorentz-invariant theory in the continuum since we want the mode spectrum to have $E \to -E$ symmetry.}
\be
E_\pm(k) = \pm \sqrt{\vec{A}^2 + \vec{\cal R}^2 \cos^2 k + \vec{\cal I}^2 \sin^2 k + 2 \vec{A} \cdot \vec{\cal R} \cos k  + 2 \vec{A} \cdot \vec{\cal I} \sin k + 2 \vec{\cal I} \cdot \vec{\cal R} \sin k \cos k}
\ee
To get a dispersion relation that is even in $k$, we require $\vec{\cal R} \cdot \vec{A} = \vec{\cal I} \cdot \vec{A} = 0$. In this case, the dispersion relation for small $k$ is
\be
E_\pm(k) = \pm \sqrt{(\vec{A} + \vec{\cal R})^2 + (\vec{\cal I}^2 - \vec{\cal R} \cdot (\vec{A} + \vec{\cal R}))k^2 + \dots}
\ee

In order to describe massless Dirac fermions in the long-wavelength limit, we want $(\vec{A} + \vec{\cal R}) = 0$ and $\vec{\cal I}^2=1$. However, we note that the energy for $k=\pi$ is $\pm |\vec{A} - \vec{\cal R}|$, so to avoid an additional massless field in the continuum limit, we want $\vec{A} - \vec{\cal R} \ne 0$ so $\vec{A} = -\vec{\cal R}$ needs to be nonzero. Without loss of generality, we can choose $\vec{\cal I} = (0,1,0)$ and $\vec{A} = -\vec{\cal R} = (0,0,t_0)$. Models with different nonzero $t_0$ are distinct but have the same continuum limit (where the relevant momenta will scale as $k \sim 1/N$ so that the higher order terms in the dispersion relation become unimportant). We will make the choice $t_0 = 1/2$. This gives
\be
H_{\text{massless}} = {1 \over 2} c^\dagger_n \sigma^z c_n - {1 \over 4} (c^\dagger_{n+1} \sigma^z c_n + c^\dagger_{n} \sigma^z c_{n+1}) + {1 \over 2} (c^\dagger_{n+1} i\sigma^y c_n -  c^\dagger_{n} i\sigma^y c_{n+1}) \; .
\ee
To describe massive fermions, we need a non-zero energy for zero momentum. Starting from the massless model, we can perturb $A_1$, $A_3$, ${\cal R}_1$, or ${\cal R}_3$ by a small parameter $\epsilon$ that we will scale as $m/N$ in the continuum limit to give the massive fermion dispersion relation for momenta $k \sim p/N$.\footnote{We could adjust ${\cal I}$ so that the coefficient of $k^2$ remains 1 in the lattice model, but this is not necessary since this coefficient will go to 1 in the continuum limit.}
We choose a model with $A_3 = 1/2 + \sin^2 q$, $A_1 = \sin q$, ${\cal R}_3 = -1/2$, and ${\cal I}_2 = 1$. We have denoted our small parameter as $\epsilon = \sin q$ and added the $\sin^2 q$ term to $A_3$ (which has no effect in the continuum limit), so that the resulting dispersion relation takes the form
\be
E_\pm(k) = \pm \sqrt{\sin^2 q + \sin^2 k + \left(\sin^2 {q \over 2} + \sin^2 {k \over 2}\right)^2}
\ee
with $k \leftrightarrow q$ symmetry. The model with these parameters also arises from a two-dimensional lattice model, where $q$ is the lattice momentum in the transverse direction.

The final lattice Hamiltonian is
\be
H(k) = \left({1 \over 2} + \sin^2 q \right) c^\dagger_n \sigma^z c_n + \sin q \; c^\dagger_n \sigma^x c_n - {1 \over 4} (c^\dagger_{n+1} \sigma^z c_n + c^\dagger_{n} \sigma^z c_{n+1}) + {1 \over 2} (c^\dagger_{n+1} i\sigma^y c_n -  c^\dagger_{n} i\sigma^y c_{n+1}).
\ee

\subsection{Continuum limit}

To define the continuum limit from the lattice model, we define a lattice spacing $a$. For a periodic chain with $N$ sites, we have length $L = Na$. In the continuum model with periodic boundary conditions, we have modes with momenta $p = 2 \pi n/L$ having energy $E = \sqrt{m^2 + p^2}$. For the lattice model, we have modes with $k = 2 \pi n/N$, so we can identify $k = p L/N = p a$. In terms of $p$, the energy in the $q \ll 1$ lattice model for a mode with $k \ll 1$ is $E \approx \sqrt{q^2 + p^2 a^2}$, so to match the continuum theory with mass $m$, we want to take $q = m a$ and rescale the Hamiltonian $H \to H/a$.

\subsection{Casimir energies for periodic and antiperiodic chains}

As a check, we can make use of the lattice model to compute Casimir energies for the 1D massless or massive Dirac fermion on a circle with periodic or antiperiodic boundary conditions.

\begin{figure}
  \centering
  \includegraphics[scale=0.5]{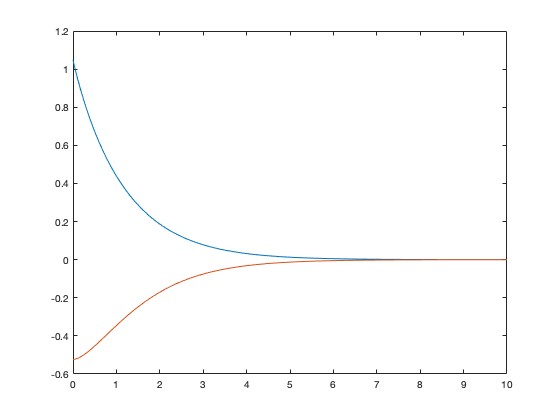}
  \caption{Casimir energy density times $L^2$ versus $m L$ for a one-dimensional Dirac fermion with mass $m$ on a circle with length $L$ with periodic (blue curve, positive function) or antiperiodic (orange curve, negative function) boundary conditions for fermions. Values for $m=0$ are $\pi/3$ and $-\pi/6$, respectively.}
\label{fig:casimirs}
\end{figure}

In the ground state for the model, all modes with negative energies are occupied (forming the Dirac sea) while all modes with positive energies are unoccupied. Thus, the ground state energies in the lattice model are
\be
E_{\text{p}}(q,N) = \sum_{k = 2 \pi n/N \in (-\pi, \pi]} -\sqrt{\sin^2 q + \sin^2 k + \left(\sin^2 {q \over 2} + \sin^2 {k \over 2}\right)^2}
\ee
for periodic boundary conditions and
\be
E_{\text{ap}}(q,N) = \sum_{k = 2 \pi (n+ 1/2)/N \in (-\pi, \pi]} -\sqrt{\sin^2 q + \sin^2 k + \left(\sin^2 {q \over 2} + \sin^2 {k \over 2}\right)^2}
\ee
for antiperiodic boundary conditions. Both of these go to minus infinity for large $N$. To extract the Casimir energies, it is convenient to define a regularized per-site energy where we subtract the per-site energy for the infinite lattice (defined using the $N \to \infty$ limit of either the periodic or antiperiodic chain).
\beas
\Delta E^{\text{site}}_{\text{p},\text{ap}}(q,N) = E_{\text{p},\text{ap}}(q,N)/N - \lim_{N \to \infty} E_{\text{p}}(q,N)/N \; .
\eeas
The Casimir energy density is then the continuum limit of this per-site energy (rescaled by $1/a = N/L$ as explained above) times the number of sites per length (which gives an additional factor of $N/L$). Thus, we have
\be
\rho^{\text{Cas}}_{\text{p},\text{ap}}(m) = {1 \over L^2} \lim_{N \to \infty} N^2 \Delta E^{\text{site}}_{\text{p},\text{ap}}(mL/N,N) \; .
\ee
The results for the dimensionless functions
\be
f_{\text{p},\text{ap}}(\mu) = L^2 \rho^{\text{Cas}}_{\text{p},\text{ap}}(\mu/L) \; ,
\ee
are plotted in Figure \ref{fig:casimirs}. In particular, the $m = 0$ results for the periodic and antiperiodic cases match with the known results
\be
f_{\text{p}}(0) = {\pi \over 3} \qquad \qquad f_{\text{ap}}(0) = - {\pi \over 6}
\ee
and in both cases, the Casimir energies fall off exponentially with mass.

We can use the results for finite mass to calculate the Casimir energy density for higher-dimensional massless Dirac fermions, since the contribution of a mode with momentum $\vec{k}$ is equivalent to the contribution of a 1D Dirac fermion with mass $|\vec{k}|$.

We find that\footnote{The normalization can be fixed here by starting with periodic transverse directions such that the energy is a sum over the contribution of each discrete transverse momentum mode. Taking the size of the periodic directions to infinity, the sums give the integral expressions here.}
\be
f^{2D}_{\text{p},\text{ap}} = \int {dk \over 2 \pi} f_{\text{p},\text{ap}}(|k|) = \int_0^\infty {dk \over \pi} f_{\text{p},\text{ap}}(k)
\ee
and
\be
f^{3D}_{\text{p},\text{ap}} = 2\int {d^2 k \over (2 \pi)^2} f_{\text{p},\text{ap}}(|\vec{k}|) = \int_0^\infty {d k \over \pi} k f_{\text{p},\text{ap}}(k)
\ee
where the factor of two comes because we have four-component spinors in 3+1 dimensions. We find good agreement with the known results $f^{2D}_{\text{p}} = \zeta(3)/\pi$, $f^{2D}_{\text{ap}} = -3\zeta(3)/(4 \pi)$, $f^{3D}_{\text{p}} = 2\pi^2/45$, $f^{3D}_{\text{ap}} = -7 \pi^2/180$ (previously obtained on the lattice in \cite{ishikawa_lat_casimir_1d_2020,ishikawa_lat_casimir_2d3d_2021,mandlecha_casimir_bag_2022}).

In the next section, we consider possible energy distributions that can be obtained via general states of the same lattice model without specifying boundary conditions.

\section{Allowed energy distributions in the lattice model}

For the lattice model with $N$ sites, we will define a local energy density operator $H_a$ associated to a site to be the on-site Hamiltonian for that site plus half of each of the terms in the Hamiltonian linking that site to its neighbors. This can be defined for sites $2 \dots (N-1)$. Each possible state of the lattice model gives some distribution of energies $E_a = \langle H_a \rangle$. We would like to ask which distributions $(E_2,\dots,E_{N-1})$ are possible, and specifically what is the lowest possible value of $E$ for which we can have $(E_2,\dots,E_{N-1}) = (E,E, \dots,E)$.

In order to answer this, it is useful to note that if any state with a distribution $(E_2,\dots,E_{n-1})$ exists, then there will also be some state of maximum von Neumann entropy with this distribution of energies. States that maximize entropy subject to the constraint that a certain set of operators $H_a$ have particular expectation values $E_n$ take the form
\be
\label{betastate}
\rho =  e^{-\sum_a \beta_a H_a}/Z(\vec{\beta}) \qquad \qquad Z(\vec{\beta}) = \tr( e^{-\sum_a \beta_a H_a}) \; .
\ee
This follows immediately from demanding a vanishing variation of the quantity $-\tr(\rho \log \rho) - \sum_a \beta_a (\tr(\rho H_a) - E_a) - \Lambda(\tr(\rho) - 1)$ with Lagrange multipliers $\beta_a$ and $\Lambda$ enforcing the constraints and the normalization condition. Thus, to understand whether a certain energy distribution is possible, we need only ask whether it is possible in the space of states of the form (\ref{betastate}).\footnote{Here, we also allow limits of sequences of such states.}

Since $H_a$ represent local operators on the lattice, the class of states we are considering have local modular Hamiltonians $K = -\log(\rho)$ built as a linear combination of the local energy density operators. In the continuum limit, this (at least naively) gives states with modular Hamitonian of the form
\be
K = \int_I dx \beta(x) T_{00}(x) \; .
\ee
As an example, for conformal field theories, the modular Hamiltonian describing an interval $[-L/2,L/2]$ for the Minkowski space vacuum is of this form, with $\beta(x) = 2 \pi ((L/2)^2 - x^2)/L$. 

\subsection{Computing the energies}
\label{sec:energies}

For free-particle Hamiltonians of the type we are considering, we can give an explicit procedure for calculating the local energies in terms of the parameters $\beta_a$.

For the general state (\ref{betastate}), each local Hamiltonian term takes the form
\be
H_a = \sum_{i,j} M^a_{i,j} c^\dagger_i c_j \; ,
\ee
where $i$ runs over the $2N$ possibilities for the site index and the component index. Suppose that the eigenvalues of $M(\vec{\beta}) = \sum_a \beta_a M^a_{i,j}$ are $\lambda_n(\vec{\beta})$ with associated normalized eigenvector $v_n$. Then we can write
\be
\sum_a \beta_a \sum_{i,j} M^a_{i,j} c^\dagger_i c_j  = \sum_n \lambda_n b^\dagger_n b_n
\ee
for some redefined creation and annihilation operators $b^\dagger_n = v_n^i c^\dagger_i$.

The partition function is then
\beas
Z &=& \tr \left( e^{- \sum_a \beta_a H_a} \right) \cr
  &=& \tr \left( e^{- \sum_n \lambda_n b^\dagger_n b_n} \right)  \cr
  &=& \prod_n (1 + e^{-\lambda_n}) \;.
\eeas
Defining
\beas
h_0(\lambda) &=& -\ln(1 + e^{- \lambda}) = {\lambda \over 2} - \ln(2 \cosh{\lambda \over 2}) \cr
h_1(\lambda) &=& h_0'(\lambda) = {1 \over 2}(1 - \tanh{\lambda \over 2}), \cr
\eeas
we have that
\be
-\ln(Z) = \sum_n h_0(\lambda_n) 
\ee
and the expectation value of $H_a$ in the state $\rho$ is
\be
E_a = {d \over d \beta_a} (- \ln(Z)) = \sum_n h_1(\lambda_n) {d \lambda_n \over d \beta_a}.
\ee
Here, using standard quantum mechanical perturbation theory (applied to the ``Hamiltonian'' $M(\beta)$), we have
\be
{d \lambda_n \over d \beta_a} = M^a_{nn}  \; ,
\ee
where we define the matrix elements
\be
M^a_{mn} \equiv v_m^\dagger M^a v_n \; .
\ee
Using these results, it is straightforward to implement an algorithm to compute $\vec{E}$ in terms of $\vec{\beta}$.

Later, it will also be useful to compute the von Neumann entropy of the state. Using the standard expression $S = -\tr(\rho \ln \rho)$, and the form (\ref{betastate}) for $\rho$, we have
\bea
S &=& \sum_a \beta_a E_a + \ln(Z) \cr
&=& \sum_{n,a} h_1(\lambda_n) \beta_a {d \lambda_n \over d \beta_a} - \sum_n h_0(\lambda_n) \; .
\label{entropy}
\eea

\subsection{Allowed energy distributions}

To formally specify the space of allowed energies, we first define the functions
\be
C(\vec{\beta}, \vec{E}) = \sum_n {1 \over 2} (\tr(\rho(\vec{\beta}) H_n) - E_n)^2 \;
\ee
and
\be
C_{\text{min}}(\vec{E}) = \min_{\vec{\beta}} C(\vec{\beta}, \vec{E}) \; .
\ee
Then the allowed energy distributions are the sets $\vec{E}$ for which $C_{min}$ vanishes.

In order to establish numerically whether a given energy distribution is allowed, we can minimize the function $C$ numerically over the space of $\beta$s, for example using a gradient descent algorithm or Newton algorithm, reviewed in Appendix \ref{sec:technical}. The gradient $\partial E_a / \partial \beta_b$ and the Hessian $\partial^2 E_a / \partial \beta_b \partial \beta_c$ used in these algorithms are computed explicitly in Appendix \ref{sec:derivatives}.

At a point in the interior of the allowed region of energies, any infinitesimal change in energies can be arranged by some infinitesimal change in $\beta$s. Since $\vec{\beta}$ and $\vec{E}$ have the same dimension, the map $\delta \vec{\beta} \to \delta \vec{E}$ is thus nonsingular and the matrix $\partial E_a / \partial \beta_b$ is invertible.

At the boundary of the region, there is some infinitesimal change in energies for which no infinitesimal change in $\beta$s will achieve, so the map from $\delta \vec{\beta} \to \delta \vec{E}$, or the matrix $\partial E_a / \partial \beta_b$ is singular. Thus, on the boundary of the allowed region of energies, we have a vanishing determinant
\be
\left| {\partial E_a \over \partial \beta_b} \right| = 0 \; .
\ee

\subsection{Minimum uniform energy density}

We now specialize to the case where all the energies $E_a$ are equal to some value $E$ and ask what is the minimum value of $E$. We find numerically that if we fix the middle $\beta_a$ (or the middle two values for even $N$), uniform energies can be achieved in a unique way by a particular set of $\beta$s that decrease from these middle values as we move toward the edges of the lattice. Within this space of uniform-energy states, the energy $E$ is found to be an decreasing function of the middle $\beta$ value(s) that exponentially approaches the minimum value as $\beta$ is increased.

In order to find this minimum value numerically, we can choose some large $\beta$ for the middle site(s)\footnote{Because of the exponential convergence of $E(\beta_{mid})$ results are essentially independent of $\beta_{mid}$ for sufficiently large $\beta_{mid}$.} and then vary the rest of the $\beta$s to minimize a cost function
\be
\label{defC2}
C_2 (\beta) = \sum_{a = 2}^{N/2-1} {1 \over 2} (E_a(\beta) - E_{\text{mid}}(\beta))^2 \; .
\ee
We find that this can be done very efficiently using Newton's method provided that the initial $\beta$s are chosen relatively close to the minimizing value. Gradient descent can be used to find these relatively close values.\footnote{It is possible to prove that the cost function has only one local minimum \cite{seraphim}, so the gradient descent will always bring us to the global minimum of the cost function where $C_2 = 0$.}

As is typical in discussions of Casimir energy, it will be convenient to normalize our energy values by subtracting the per-site energy for the $N \to \infty$ limit of a periodic chain. Thus, we define a subtracted minimum per-site energy
\be
\Delta E_{\text{min}}(q,N) = E_{\text{min}}(q,N) - E_{\text{p}}(q,\infty) \; .
\ee

\subsection{Continuum limit}

We can use these lattice results to define a function $f_{\text{min}}(\mu)$ that gives the minimum energy density (in units of $1/L$) for a 1D Dirac fermion of mass $\mu/L$ on an open interval of width $L$ for states where the energy density is uniform in the interval. We define (based on the scaling explained in section 2.3)
\be
f_{\text{min}}(\mu,N) = N^2 \Delta E_{\text{min}}(\mu/N,N) \; .
\ee
and
\be
f_{\text{min}}(\mu) = \lim_{N \to \infty} f_{\text{min}}(\mu,N) \; .
\ee
We will compare this to the results $f_{\text{p}},f_{\text{a}}$ defined in section \ref{sec:setup} giving the energy density for the periodic and antiperiodic chains.

For fermions in 2+1 or 3+1 dimensions, we can construct states with uniform energy density by placing the degrees of freedom associated with transverse momentum $\vec{p}$ in the uniform-energy state with minimum energy for mass $|\vec{p}|$. In this case, the energy density for the 2+1 dimensional case\footnote{It is possible that there are states with even lower uniform energy density such that the energy density in the individual transverse momentum modes is not uniform on the interval but the overall energy density is. Our results should thus be seen as an upper bound on the minimum possible uniform energy density.}
\be
\label{f2D}
f^{2D}_{\text{min}} = \int {dp \over 2 \pi} f_{\text{min}}(|p|) = \int_0^\infty {dp \over \pi} f_{min}(p)
\ee
and
\be
\label{f3D}
f^{3D}_{\text{min}} = 2\int {d^2 p \over (2 \pi)^2} f_{\text{min}}(|\vec{p}|) = \int_0^\infty {d p \over \pi} p f_{\text{min}}(p) \; .
\ee

A crucial point will be to understand the asymptotic behaviour of $f_{\text{min}}(p)$ for large $p$ to see whether these integrals converge. If not, uniform negative energy densities with arbitrarily large magnitude are possible for any fixed interval width.

\section{Results}

Implementing the lattice calculations using MatLab, we have generated results for $f_{\text{min}}(\mu,N)$ up to $N=100$. In each case, the Newton method algorithm allows us to find $\beta_a$s such that the energies are equal to any desired accuracy, and we have checked to ensure that the central $\beta$ values are chosen large enough so that increasing them produces no significant change in the energies.

\begin{figure}
  \centering
  \includegraphics[scale=0.38]{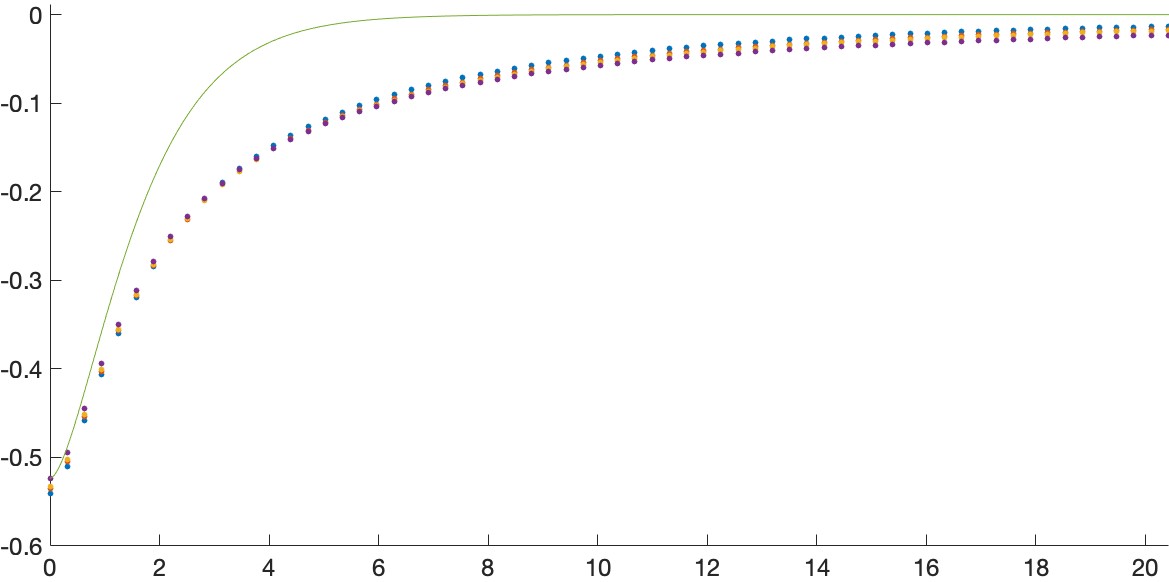}
  \caption{Numerical results for $f_{\text{min}}(mL,N)$  (lattice approximation to $L^2 \Delta E_{\text{min}}$) vs $mL$ for $N = 30$ (blue), $50$ (red), $70$ (orange), and extrapolated to $\infty$ (purple). The solid green line shows the Casimir energy density for antiperiodic boundary conditions.}
\label{fig:dataAP}
\end{figure}

Our numerical results for $f_{\text{min}}(\mu,N)$ vs $\mu$ are plotted for various values of $N$ in Figure \ref{fig:dataAP} along with the antiperiodic energy density $f_{\text{a}}(\mu)$. For each $\mu$ the results appear to converge as expected to a limit $f_{\text{min}}(\mu)$ that we interpret as $L^2$ times the minimum uniform energy density for a 1+1 dimensional Dirac fermion of mass $\mu / L$.

\subsubsection*{Minimum energy density: massless $1+1$D Dirac fermion}

\begin{figure}
  \centering
  \includegraphics[scale=0.45]{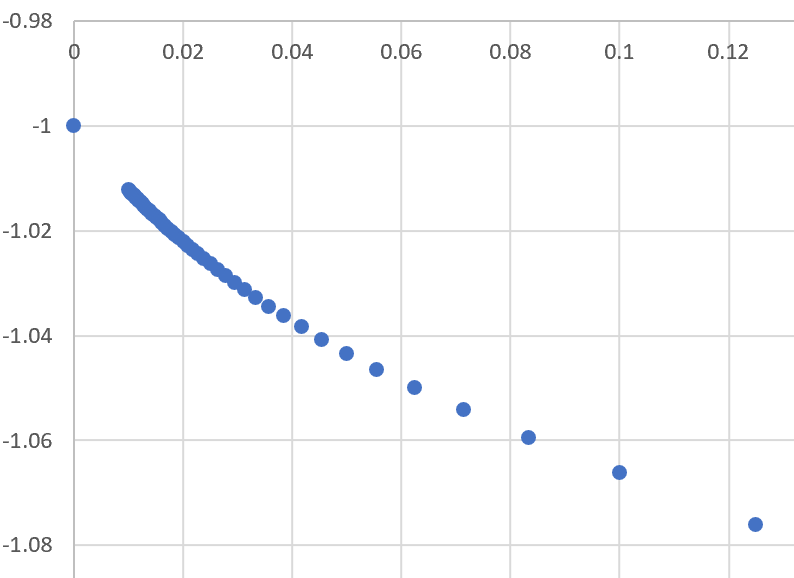}
  \caption{Numerical results for $f_{\text{min}}(\mu=0,N)/|f_{\text{a}}(\mu = 0, N = \infty)|$ vs $1/N$. Value shown on the vertical axis corresponds to the continuum result for antiperiodic boundary conditions.}
\label{fig:zerodat}
\end{figure}

We see that for $\mu = 0$, the results appear to converge to the antiperiodic energy $f_a(0)$. The $\mu = 0$ results for $f_{\text{min}}(\mu,N)/f_a$ are plotted vs $1/N$ in Figure \ref{fig:zerodat}. The data appear to approach 1 with a finite slope, indicating that $f_{min}(\mu,N)/f_a = 1 + c_1/N + c_2/N^2 + \dots$. Fitting the data for $N=60$ to $N=100$ to a fifth order polynomial in $1/N$, the constant term gives $1.000071$, so the result for the minimum possible uniform energy density of a 1D Dirac fermion on an interval matches with the ground state energy density with antiperiodic boundary conditions on the interval to an accuracy of $0.01\%$.\footnote{The result doesn't change significantly if we vary the order of the polynomial or the number of points used in the fit. Fitting a quadratic polynomial to the three points with $N=96,98,100$ already gives $0.1\%$ accuracy.} Thus, the numerical results suggest that antiperiodic boundary conditions provide the least energy among all possible ways of extending the physical system past the boundaries of the open interval. Motivated by this observation, we have been able to prove this fact directly via the Quantum Energy Inequalities (see Appendix D).

\subsubsection*{Minimum energy density: massive $1+1$D Dirac fermion}

\begin{figure}
\centering
  \includegraphics[scale=0.60]{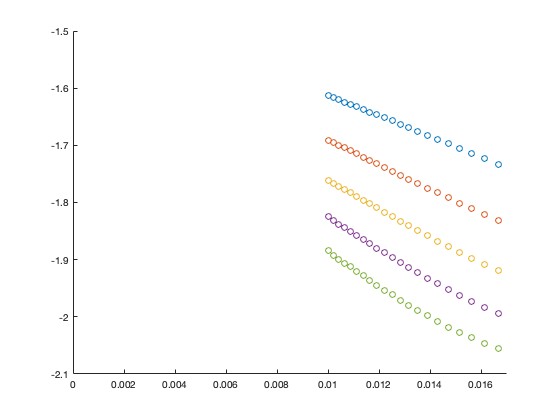}
  \caption{The function $n(N,\mu)$ vs $1/N$ for $\mu = 6 \pi, 7 \pi, 8 \pi, 9 \pi, 10\pi$ (top to bottom). This function should approach the exponent of the power law fall-off of $f_{min}(\mu)$ vs $\mu$ for large $N$ and large $\mu$. Quadratic extrapolations of each curve to $1/N = 0$ all fall within the range $-1.50 \pm 0.05$.}
\label{fig:powerextr}
\end{figure}

For any non-zero mass, Figure \ref{fig:dataAP} suggests that the minimum energy density is strictly less than that for antiperiodic boundary conditions (i.e has a greater magnitude). While $f_a(\mu)$ falls off exponentially with $\mu$, our numerical results for $f_{\text{min}}(\mu,N)$ appear to extrapolate to a function $f_{\text{min}}(\mu)$ that falls off only as a power of $\mu$. In order to estimate this power, we calculate 
\be
n(N,\mu) = {\mu \over f_{\text{min}}} {d f_{\text{min}} \over d \mu} 
\ee
For a power law behavior $f \sim \mu^{n_0}$, this function should approach the power $n_0$ for large $N$ and large $\mu$. The results for $n(N,\mu)$ vs $1/N$ for $\mu = 6 \pi,7\pi,8\pi,9\pi$ are plotted in Figure \ref{fig:powerextr}. In each case, a quadratic extrapolation of the curve gives a result in the range $-1.50 \pm 0.05$, consistent with a power law fall off  $f_{\text{min}}(\mu) \sim \mu^{-3/2}$.

\subsubsection*{Minimum energy density: massless $2+1$D Dirac fermion}

\begin{figure}
\centering
  \includegraphics[scale=0.70]{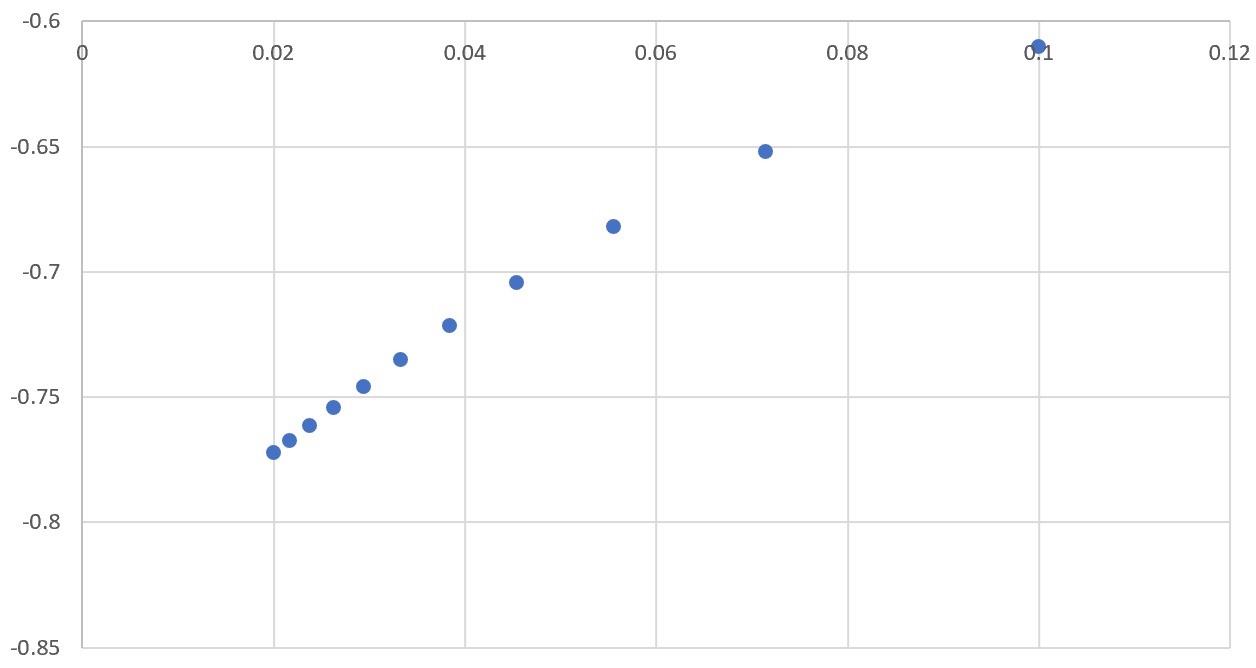}
  \caption{Lattice approximation to $f^{2D}_{\text{min}}$ vs inverse lattice size $1/N$.}
\label{fig:int2D}
\end{figure}

For the observed falloff in $f_{\text{min}}(\mu)$, the result (\ref{f2D}) for the minimum energy density in a 2D Dirac fermion (with uniform energy density in each mode) converges. Our numerical results for the integral for various $N$ are shown in Figure \ref{fig:int2D}. The result appears to extrapolate to a value of $f^{2D}_{\text{min}} \approx -0.83$ in the limit $1/N = 0$. This is a larger in magnitude than the energy density $f^{2D}_{\text{ap}} = -3 \zeta(3)/4 \pi$ with antiperiodic boundary conditions for fermions by a factor of close to three.

\subsubsection*{Minimum energy density: massless $3+1$D Dirac fermion}

\begin{figure}
\centering
  \includegraphics[scale=0.65
  ]{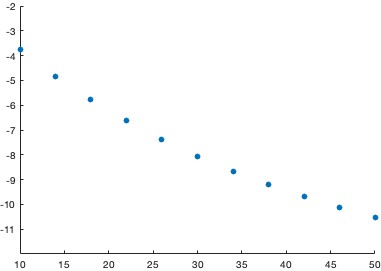}
  \caption{Estimate for $f^{3D}_{min}$ versus lattice size $N$.}
\label{fig:int3D}
\end{figure}

In order for the integral in (\ref{f3D}) to converge, $f_{\text{min}}(\mu)$ would have to fall off faster than $1/\mu^2$ for large $\mu$. Our data suggest that this is not the case. Thus, our results suggest that there are states of a 3+1 dimensional Dirac fermion field on an open interval of length $L$ with arbitrarily small uniform energy density. Our numerical results for the integral for various $N$ are shown in Figure \ref{fig:int3D}.

\subsection{Behavior of $\beta(x)$ for various $m$.}

\begin{figure}
  \includegraphics[scale=0.4]{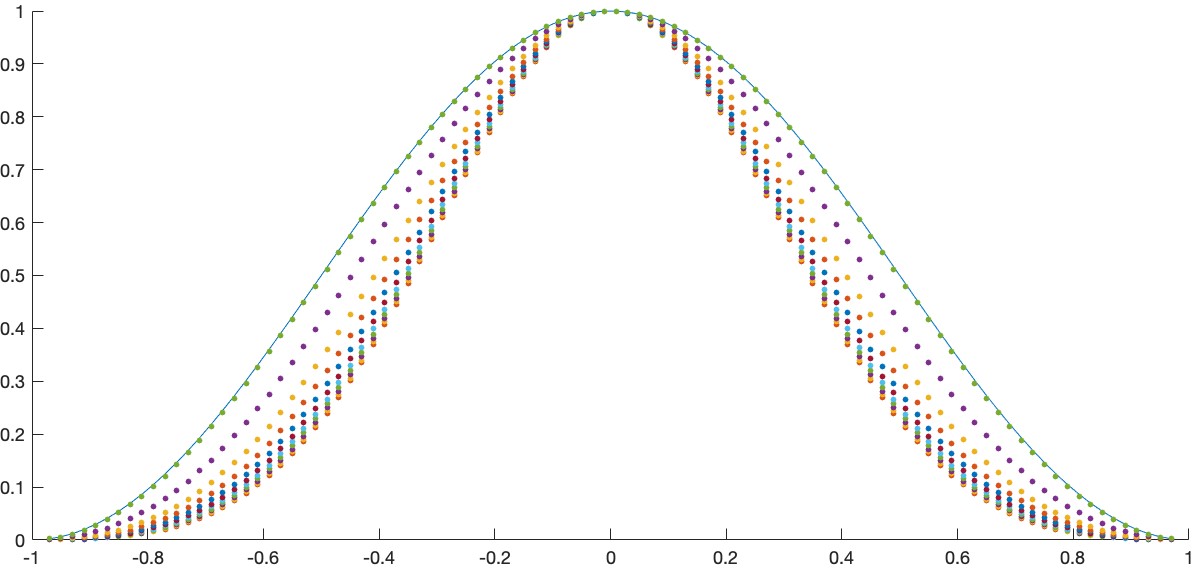}
  \caption{Distribution of $\beta_a$ values (normalized) for $N=100$ and $\mu = 0,\pi,\dots, 10 \pi$ (top to bottom). Smooth curve corresponds to continuum $\hat{\beta}(x)$ in the antiperiodic vacuum of the massless fermion.}
\label{fig:beta100}
\end{figure}

It is interesting to look at the behavior of the $\beta$ parameters giving states that minimize the uniform energy densities. We find that the profile of $\beta$s appears to approach a particular shape $\beta(x)$ for each $\mu$.
In Figure \ref{fig:beta100}, we plot the distribution of $\beta$s for $N=100$ at various $\mu$. We see that in each case, the behavior of $\beta(x)$ near the boundaries appears to be a function that increases quadratically from zero.\footnote{Fitting the first few points for a given $N$ to a quadratic function, we find that the minimum of this quadratic function appears to approach 0 as $N$ increases.}

We can compare our results for the massless case to the expected results for the modular Hamiltonian of the reduced state on an interval where we start from the ground state of a fermion with antiperiodic boundary conditions. Setting the size of the periodic direction to $L=1$, the modular Hamiltonian for fermions on an interval $[-R,R]$ was shown in \cite{cardy_mod_ham_cft2_2016} to correspond to
\be
\beta_R(x) = 2\pi {\sin(\pi(R-x)) \sin(\pi (R+x)) \over \pi \sin (2 \pi R)}.
\ee
In the limit $R \to 1/2$ where the interval covers the entire circle, all values of $\beta(x)$ approach infinity. Defining $\hat{\beta}(x) = \beta(x)/\beta(0)$, we find that for $R \to 1/2$ we get a limiting value
\be
\hat{\beta}(x) = \cos^2(\pi x)
\ee
This agrees well with our results for $\hat{\beta}(x)$ at $m=0$, as shown in Figure \ref{fig:beta100}.

An interesting point is that for any $R \in (0, 1/2)$, $\beta_R(x)$ increases linearly with $R-x$ with slope $2\pi$ as we move inward from the boundary of the interval. But the coefficient of the quadratic correction to this diverges as $R \to 1/2$. In the $R \to 1/2$ limit of the normalized function $\hat{\beta}$, it is only this quadratic piece that remains.

\subsection{Other components of the stress tensor}
\label{sec:Tzz}

In our investigation, we have searched for the states with the lowest possible uniform value of the energy density $T_{00}$. It is also interesting to look at other components of the stress-energy tensor for these states. For Poincar\'e-invariant ground states of a massless Dirac fermion, we have seen above that the entire stress-energy tensor is spatially constant, with $T_{zz} = T_{00}$.

To investigate the behavior of $T_{zz}$ using the lattice model, we note that in the continuum theory the on-shell stress tensor component $T_{zz}$ for a massive 1+1 dimensional Dirac fermion can be obtained by setting $m=0$ in the expression for $T_{00}$. Thus, we can take the lattice version of $T_{zz}(x_i) a$ as $H_i(m=0)$. This guarantees that the tracelessness condition $\langle T_{00} \rangle = \langle T_{zz} \rangle$ holds in the lattice theory for the massless case also. 

\begin{figure}
  \includegraphics[scale=0.6]{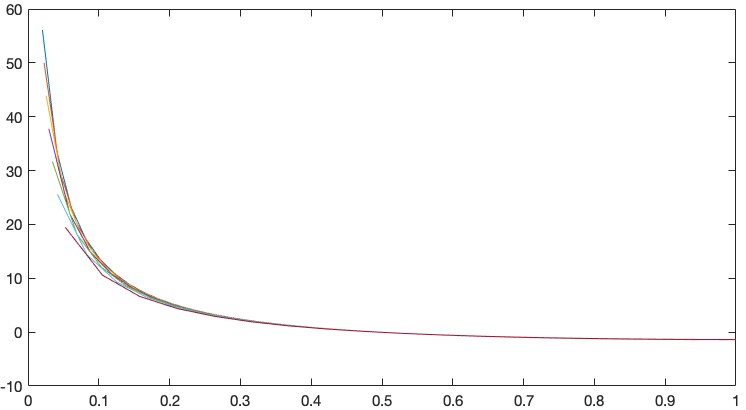}
  \caption{$T_{zz} L^2$ vs $z$ for the half-interval $[-L/2,0]$ in states of the $m = \pi /L$ 1+1 dimensional Dirac fermion with minimal uniform energy densities for $N=40,50,\dots,100$} (bottom to top).
\label{fig:Tzz}
\end{figure}

For our uniform energy-minimizing states with non-zero mass, we find that the $T_{zz}$ is generally non-uniform, starting from a negative value at the middle of the interval and becoming large and positive towards the edges. The example of $p = \pi/L$ is shown in Figure \ref{fig:Tzz}. 

We note that the null energy $T_{00} + T_{zz}$ is negative for a large portion of the interval. It would be interesting to investigate states constrained to have uniform null energy or both $T_{00}$ and $T_{zz}$ uniform, but we leave this for future work. 

\subsection{Entropy}

It is interesting to consider the entropy for the states we have been considering. For a fixed $N$ and parameters $\beta_a$, the entropy can be computed using (\ref{entropy}). 

\subsubsection*{Fixed $N$, $\beta_{\text{mid}} \to \infty$}
We find that keeping $N$ fixed and taking $\beta_{\text{mid}}$ (the $\beta$ value for the middle site(s)) to infinity, the entropy goes to zero. We can understand this as follows:
our energy-minimizing states are the ground states for certain lattice Hamiltonians that break translation symmetry. We recall that the density matrix takes the form 
\be
\rho =  e^{-\sum_a \beta_a H_a}/Z,
\ee
where the $\beta_{\text{mid}}$ corresponding to the central site(s) are taken to infinity, and the $\hat{\beta}_a = \beta_a/\beta_{\text{mid}}$ for the remaining sites are adjusted to give a uniform energy density. We can rewrite this as
\be
\rho = \lim_{\beta \to \infty} e^{-\beta \hat{H}}/Z,
\ee
where $\hat{H} = \sum_a \hat{\beta}_a^\infty H_a$,
so the limiting state is the ground state of $\hat{H}$. 

\subsubsection*{Continuum field theory subsystems}

The previous results may appear to be in tension with the idea that the entropy of quantum field theory subsystems should be divergent in the continuum limit. For example, the entropy of a CFT subsystem of length $L$ should be $c/3 \log (L / \epsilon)$ in the vacuum state and have a similar $\log(1/\epsilon)$ divergence in any other state. This should translate to a $\log(N)$ behavior for our lattice regularization.

In order to see this behavior, we need to be careful about how the $\beta$ parameters are taken to infinity. For a continuum state with modular Hamiltonian
\be
K = \int dx \beta(x) T_{00}(x)
\ee
we can write the discretized version with lattice scale $a$ as
\be
\sum_i \beta(x_i) (a T_{00}(x_i))
\ee
The expression in brackets is the energy associated with a site; we have argued above that this becomes
\be
a T_{00}(x_i) \sim {1 \over a} H_i \; .
\ee
Thus, the lattice equivalent of the modular Hamiltonian above is (using $a = L/N$)
\be
K_{\text{discrete}} = \sum_i {N \beta(x_i) \over L} H_i \equiv \sum_i \beta_i H_i
\ee
In order to faithfully represent the continuum QFT state, we want the lattice $\beta$s to scale with $N$ as we take $N \to \infty$, $\beta_i = N/L \beta(x_i)$. Taking $N$ to infinity with the $\beta$ parameters scaling in this way, we can have the expected $\log(N)$ behavior of the entropy. 

\subsection{Aside: Lattice ground states}
\label{sec:latgnd}

\begin{figure}
  \centering
  \includegraphics[scale=0.6]{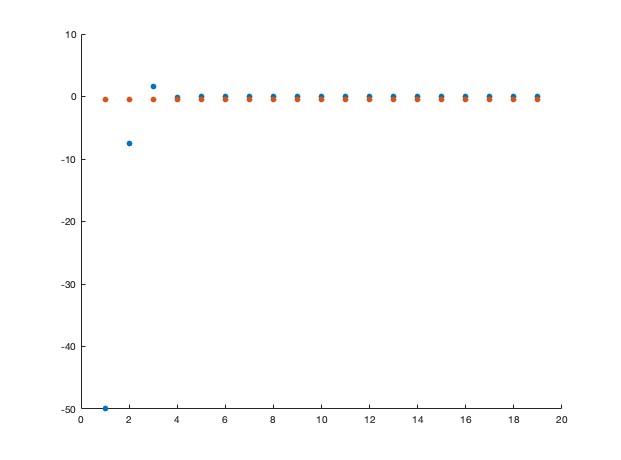}
  \caption{Lattice energies for sites $1 \dots 19$ in the $N=40$ lattice model for the massless Dirac fermion. Blue points show the local energies in the lattice ground state (relative to the energies in the infinite chain); orange points show the energies for the minmum-energy state with uniform energies.}
\label{fig:truemin}
\end{figure}

As an aside, we note that the minimum energy states with a uniform energy distribution are not the states with minimum total energy for the lattice. The actual minimum energy states are obtained by taking all $\beta$s to infinity, or alternatively by occupying all negative energy modes. The minimum energy is then just the sum of the negative eigenvalues of $H$. For these states, we find that the energy relative to the periodic lattice goes to $-\infty$ after the continuum scaling. However, most of this negative energy lives at the outermost sites. In the $N \to \infty$ limit, we find that the energy for site $n$ (before rescaling $H \to H/a$) approaches a limit $E_n(q)$. For example, in lattice for the massless theory, the subtracted lattice energies at large $N$ (starting from the outermost site) are $-0.03112, -0.00472, 0.00097, -0.00011, \dots$ decaying quickly to zero. After the scaling $N^2 \Delta E(\mu/N,N)$ appropriate to define the continuum limit, we find an infinite energy localized to the boundary of the interval for $N \to \infty$ with a finite energy density inside. The lattice energy for the interior points in the state with least energy is not as low as that for the energy-minimizing state with uniform energy, as we see in Figure \ref{fig:truemin}. To be clear, both of the states shown in Figure \ref{fig:truemin} are allowed physical states, one would arise as the ground state for the chain with antiperiodic boundary conditions, while the other would arise as the ground state with some other boundary conditions where the two ends of the chain do not interact.

\section{Discussion}

For massless Dirac fermions in 1+1, 2+1, and 3+1 dimensions, we have explored the allowed values of energy density (relative to the Minkowski vacuum) in situations where this energy density is uniform in a slab of width $L$ in one of the spatial directions. 

For 1+1 dimensions, our numerical results suggest that this energy must always be larger than or equal to the value in the situation where the two sides of the slab are identified with antiperiodic boundary conditions for fermions and we take the vacuum state on the resulting circle. Motivated by these results, we were able to prove this using the Quantum Energy Inequalities (see Appendix D). For 2+1 dimensions, we find that the energy density can be lower than the antiperiodic vacuum, at least by some finite factor. For 3+1 dimensions, we find evidence that there is no lower bound on the energy density. 

The states that we find are apparently not vacuum states for the systems we consider. The lattice results suggest that we can lower the overall energy by considering states with non-uniform energy distributions where most of the negative energy is moved to the boundary of the slab (see Section \ref{sec:latgnd}). We have seen that $T_{zz}$ is non-uniform, so the conservation equations suggest that the energy density will evolve with time.  

It would be interesting to investigate the constraints on uniform energy density with the additional requirement that $T_{zz}$ is also uniform, or that $T_{zz} = (d-1) T_{00}$ as we must have for states that preserve $(d-1)$-dimensional Poincar\'e invariance. We could proceed in a similar way, considering states of the form 
\be
\log \rho = -\beta_i H_i(m) - \alpha_i H_i (m=0)
\ee
taking $H_i(m=0)$ as the lattice version of $T_{zz}$ as discussed in Section \ref{sec:Tzz}. 

It would be interesting to better understand what sort of physics can give rise to such states. Our data suggest that the maximum negative energy magnitude only falls off polynomially with the mass in 1+1D, whereas the usual Casimir effect approaches its large mass limit exponentially fast in $mL$. This is probably related to the fact that unlike the Casimir effect situation, our states are not ground states for some boundary conditions (otherwise, they would be time independent). 

Finally, one could search for analogous effects in bosonic systems, e.g. photons. If states with enhanced negative energy could also be found there, then one could explore their potential experimental realization. One could also ask about experimental realization of our fermion results in graphene.

\textit{Acknowledgements}---We thank Stefano Antonini, Petar Simidzija, and Chris Waddell for collaboration on related topics and Felipe Rosso for discussions. We acknowledge support from the U.S. Department of Energy grant DE-SC0009986 (B.G.S.), the National Science and Engineering Research Council of Canada (NSERC) and the Simons foundation via a Simons Investigator Award and the ``It From Qubit'' collaboration grant.

\appendix

\section{Dirac fermions}
\label{sec:fermions}

In this appendix, we review how Dirac fermions in 2+1 or 3+1 dimensions with fixed momentum in one or two transverse directions respectively are equivalent to one-dimensional Dirac fermions with mass $|\vec{k}|$.

\subsubsection*{Two plus one dimensions}

In 2+1 dimensions, we can decompose the spinor as
\be
\psi(x^\mu,y) = \int {dk \over 2 \pi} \psi_{k} e^{i k y}
\ee
In this case, the theory decomposes into decoupled momentum modes with one-dimensional Lagrangian density
\be
{\cal L}_k = i \bar{\psi}_k \gamma^\mu \partial_\mu \psi_k - k \bar{\psi}_k \gamma^2 \psi_k
\ee
Defining Dirac matrices as
\be
\gamma^0 = \sigma^x \qquad \gamma^1 = i \sigma^y \qquad \gamma^2 = i \sigma^z
\ee
we can write the action explicitly in terms of spinor components
\be
\psi_k = \left( \ba{c} \psi_{k+} \cr \psi_{k-} \ea \right)
\ee
as
\be
i \psi_{k+}^* \partial_- \psi_{k+} + i \psi_{k+}^* \partial_+ \psi_{k+} + i k (\psi_+^* \psi_- - \psi_-^* \psi_+)
\ee
A redefinition of the fields $\psi_+ \to e^{i \pi/4} \psi_+$, $\psi_- \to e^{-i \pi/4} \psi_-$ gives
\be
i \psi_{k+}^* \partial_- \psi_{k+} + i \psi_{k+}^* \partial_+ \psi_{k+} +  k (\psi_+^* \psi_- + \psi_-^* \psi_+)
\ee
which is the one-dimensional Dirac action for a fermion of mass k:
\be
{\cal L}_k = i \bar{\psi}_k \gamma^\mu \partial_\mu \psi_k + k \bar{\psi}_k \psi_k \; .
\ee

\subsubsection*{Three plus one dimensions}

In 3+1 dimensions, we have four-component spinors that we can write in terms of two component fields $\psi$ and $\chi$ as
\be
\Psi = \left( \ba{l} \psi \cr \chi \ea \right) \; .
\ee
The four-dimensional Dirac matrices can be defined as where the second factor in the tensor product is a matrix that
\be
\Gamma^0 =  \sigma^x \otimes \sigma^x \qquad \Gamma^1 = i \sigma^y \otimes \sigma^x \qquad \Gamma^2 = i\sigma^z \otimes \sigma^x \qquad \Gamma^3 = \mathbb{I} \otimes i \sigma^y \; .
\ee
In this case, writing
\be
\Psi = \left( \ba{l} \psi_{\vec{k}} \cr \chi_{\vec{k}} \ea \right) e^{i k_y y + i k_z z}
\ee
the four-dimensional action decomposes into terms for each value of the transverse momentum $\vec{k} = (k_x,k_y)$,
\beas
{\cal L}_{\vec{k}} =&&
i \bar{\psi}_{\vec{k}} \gamma^\mu \partial_\mu \psi - k_y \bar{\psi} \gamma^2 \psi + k_z \bar{\psi} \psi \cr
&&i \bar{\chi}_{\vec{k}} \gamma^\mu \partial_\mu \chi - k_y \bar{\chi} \gamma^2 \chi - k_z \bar{\chi} \chi
\eeas
By the $y-z$ rotational symmetry of the underlying higher-dimensional action, we can transform $(k_y,k_z) \to (0,\sqrt{k_y^2 + k_z^2})$. Redefining the Dirac matrices as $\gamma^\mu \to - \gamma^\mu$ for the $\chi$ terms gives finally
\beas
{\cal L}_{\vec{k}} =&&
i \bar{\psi}_{\vec{k}} \gamma^\mu \partial_\mu \psi  + |\vec{k}| \bar{\psi} \psi \cr
&&i \bar{\chi}_{\vec{k}} \gamma^\mu \partial_\mu \chi + |\vec{k}| \bar{\chi} \chi
\eeas
so we have a pair of one-dimensional Dirac fermions with mass $|\vec{k}|$.

\section{Energies and their derivatives from $\beta$s}
\label{sec:derivatives}

In applying the gradient descent and Newton algorithms to find $\vec{\beta}$ parameters that give rise to uniform energy densities, it will be useful to have explicit formulae for the first and second derivatives of $E_a$ with respect to $\beta_b$. Making use of the definitions in Section \ref{sec:energies}, along with
 \beas
h_2(\lambda) &=& h_1'(\lambda) = -{1 \over 4}(1 - \tanh^2{\lambda \over 2}) \cr
h_3(\lambda) &=& h_2'(\lambda) = {1 \over 4} \tanh{\lambda \over 2} (1 - \tanh^2{\lambda \over 2})
\eeas
we find that
\be
{d E_a \over d \beta_b} = \sum_n h_2(\lambda_n) {d \lambda_n \over d \beta_a} {d \lambda_n \over d \beta_a} + \sum_n h_1(\lambda_n) {d^2 \lambda_n \over d \beta_a d \beta_b} \; .
\ee
This is symmetric in $a$ and $b$ since it is the second derivative of $-\ln(Z)$.
From second-order quantum mechanics perturbation theory, the second derivative of the eigenvalues is\footnote{Here, it is important that $M$ is linear in the $\beta$s, otherwise, there would be a term involving the matrix element of the second derivative of $M$ with respect to betas.}
\be
{d^2 \lambda_n \over d \beta_a d \beta_b} = \sum_{m \ne n} {M^a_{nm} M^b_{mn} \over \lambda_n - \lambda_m} + {a \leftrightarrow b} \; .
\ee
Finally, it will be useful to have
\beas
{d^2 E_a \over d \beta_b d \beta_c} &=& \sum_n h_3(\lambda_n) {d \lambda_n \over d \beta_a} {d \lambda_n \over d \beta_a} {d \lambda_n \over d \beta_c} \cr && + \sum_n h_2(\lambda_n) \left( {d^2 \lambda_n \over d \beta_a d \beta_b} {d \lambda_n \over d \beta_c} + {d^2 \lambda_n \over d \beta_b d \beta_c} {d \lambda_n \over d \beta_a}  + {d^2 \lambda_n \over d \beta_c d \beta_a} {d \lambda_n \over d \beta_b} \right)  \cr
&&+ \sum_n h_1(\lambda_n) {d^3 \lambda_n \over d \beta_a d \beta_b d \beta_c} \; .
\eeas
This is totally symmetric in $a$,$b$, and $c$. To compute the third derivative, we need the third-order quantum mechanics perturbation theory result
\be
{d^3 \lambda_n \over d \beta_a d \beta_b d \beta_c} = \left(\sum_{m \ne n,l \ne n}{M^a_{nm} M^b_{ml} M^c_{l n} \over (\lambda_n - \lambda_m)(\lambda_n - \lambda_l)} - \sum_{l \ne n} M^a_{nn} {M^b_{nl} M^c_{ln} \over (\lambda_n - \lambda_l)^2}\right) +  5 \; abc \; {\rm permutations}
\ee

Now, suppose we wish to find $\beta$ parameters which give a set of energies $E_a^0$. We can consider a cost function
\be
C = \sum_a {1 \over 2} (E_a(\vec{\beta}) - E_a^0)^2 \; .
\ee
The gradient of this cost function is
\be
{d C \over d \beta_b} = \sum_a (E_a - E_a^0) {d E_a \over d \beta_b} \; .
\ee
The Hessian of this cost function is
\be
{d C \over d \beta_b d \beta_c} = \sum_a {d E_a \over d \beta_b} {d E_a \over d \beta_c} 
+ \sum_a (E_a - E_a^0) {d^2 E_a \over d \beta_b d \beta_c} \; .
\ee
Alternatively, suppose we wish to minimize a cost function
\be
C_2 = \sum_a {1 \over 2} (E_a - E_M)^2 \;.
\label{cost2}
\ee
in order to set all of the energies equal ($M$ is some specific site). The gradient of this cost function is
\be
{d C_2 \over d \beta_b} = \sum_a (E_a - E_M) \left( {d E_a \over d \beta_b} - {d E_M \over d \beta_b}\right) \; .
\ee
The Hessian of this cost function is
\be
{d C_2 \over d \beta_b d \beta_c} = \sum_a (E_a - E_M) \left( {d^2 E_a \over d \beta_b d \beta_c} - {d^2 E_M \over d \beta_b d \beta_c}\right) + \sum_a \left( {d E_a \over d \beta_b} - {d E_M \over d \beta_b}\right) \left( {d E_a \over d \beta_c} - {d E_M \over d \beta_c}\right) \; .
\ee

\section{Technical notes}
\label{sec:technical}

In generating the data for $f_{min}(\mu,N)$, we need to calculate the lattice energy $\Delta E(q=\mu N, N)$ by choosing large value for the middle $\beta$s and then adjusting the remaining $\beta$s to achieve equal energies at all sites. As described in the main text, we can do this by employing either gradient descent or the Newton method in order minimize the cost function $C_2$ defined in (\ref{cost2}). For gradient descent, we update the $\beta$ parameters in each step as
\be
\beta_a \to \beta_a - \alpha {\partial C_2 \over \partial \beta_a}
\ee
for some chosen $\alpha$. For the Newton method, in each step we approximate the cost function by a quadratic surface and then jump to the parameter values corresponding to the minimum of this surface. Explicitly, defining
\be
H_{ab} = {\partial^2 C_2 \over \partial \beta_a \partial \beta_b}
\ee
evaluated at the current point, we update the parameters as
\be
\beta_a \to \beta_a - H^{-1}_{ab} {\partial C_2 \over \partial \beta_b} \; .
\ee
The Newton method is very efficient provided that we start close enough to the minimum. Formulae for the derivatives needed in these two methods are provided in Appendix \ref{sec:derivatives}

In practice, we generate data either by fixing $\mu$ and moving to progressively larger $N$ or fixing $N$ and generating data for progressively larger $\mu$. It at each step, it is very useful to guess the $\beta$s based on those for the previous steps. For example, when generating data for some arithmetic sequence of $\mu$ values, we can guess the next value of $\vec{\beta}$ based on an $(M-1)$st order fit to the previous $M$ values, giving
\be
\vec{\beta}_n = \sum_{l=1}^M {M \choose l} (-1)^{(l-1)} \vec{\beta}_{n-l} \; .
\ee
In this way, it is usually possible to get a starting value of $\vec{\beta}$ such that only a few steps of the Newton algorithm yield $\beta$s giving almost identical energies at each site.

\section{Proof that the antiperiodic vacuum has the least uniform energy density for a 1+1 dimensional CFT on an interval of fixed length.}

Consider any state of a quantum field theory on the domain of dependence of an open interval of length $L$. Regardless of the physics outside this domain of dependence, such a state can be purified by a state of the quantum field theory on Minkowski space, since the purification involves an infinite number of quantum field theory degrees of freedom outside the interval.\footnote{We will take this as an assumption; our argument applies to quantum field theories for which this is true, though we believe it should be true generally.} In this case, if particular uniform energy density is allowed for the field theory on an interval of length $L$ with some unspecified boundary physics beyond that, the same uniform energy density over an interval of length $L$ should be possible at an instant in time for some state of the field theory on Minkowski space. 

For a 1+1 dimensional CFT in Minkowski space, the allowed stress-energy tensors are constrained by the quantum energy inequalities (QEI) \cite{Flanagan:1997gn, Vollick:2000pm,Fewster:2004nj, Fewster:2012yh}. The energy density splits into components
\be
 \langle T^{00}(x,t) \rangle =  \langle T_{++}(x^+) \rangle  +  \langle T_{--}(x^-) \rangle  
\ee
where $x^{\pm} = t \pm x$ and the two components on the right are constrained to depend only on a single lightlike coordinate because of conservation and tracelessness. For $T = \langle T_{++} \rangle $ or $\langle T_{--} \rangle$, the QEI restricts the dependence on s = $x^{\pm}$ to satisfy 
\be
\int \langle T(s) \rangle f(s) ds \ge -{c \over 12 \pi} \int \left({d \over ds} \sqrt{f(s)} \right)^2 ds \; ,
\ee
where $f$ is any non-negative function of Schwartz class
(i.e. such that the function and its derivatives vanishes at infinity faster than any inverse power). Roughly, this can be understood as arising from the requirement that the total energy of the state is positive in any conformal frame (i.e. after a general local conformal transformation). 

In order to make use of the QEIs, it is useful to express them in a different way, defining $\psi(s) = {\cal N} \sqrt{f(s)}$ where ${\cal N}$ is a normalization chosen so that $\psi^2(s)$ integrates to 1. In this case, the QEI is equivalent to the statement that
\be
\label{qeng}
\int ds \psi(s) \left( -{c \over 12 \pi} {d \over ds^2} + T(s)  \right) \psi(s) \ge 0
\ee
for all positive $\psi(s)$ of Schwartz class. This precisely the expectation value of the energy in the state described by wavefunction $\psi(s)$ for a quantum particle moving in one dimension with
potential $T(s)$. It follows almost immediately that {\it the quantum energy inequalities are satisfied if and only if the Schrodinger problem with potential $T(s)$ has no states
with negative energy}.\footnote{If the QEI are violated, then there exists some $\psi(s)$ for which (\ref{qeng}) is violated, so the state described by wavefunction $\psi(s)$ has negative energy. Conversely, suppose that the quantum system with potential $T(s)$ has a state with negative energy. Then the ground state of this system also has negative energy. But, energy eigenstate wavefunctions can be taken to be real, and the ground state wavefunction has no nodes, so there exists a square-integreble positive
real function such that (\ref{qeng}) is violated. This function can be defined as the limit of a series of functions of Schwarz-class, so the QEI are violated.}

Now suppose that $T(s)$ is uniform on $[0,L]$ with value $T_0$. We would like to ask what the lowest allowed value of $T_0$ is. To avoid negative energy states for the quantum mechanics problem with potential $T(s)$, the best we can do is make $T(s)$ as large as possible outside the interval $[0,L]$. This gives us an infinite square well with size $L$ and potential $T_0$. The ground state energy in this case (using that the usual $\hbar^2/(2m)$ is replaced here by $c/(12 \pi)$) is 
\be
E_0 = T_0 + {c \pi \over 12 \pi L^2} \; .
\ee
Requiring that this is non-negative gives
\be
\label{T0const}
T_0 \ge -{c \pi \over 12 L^2}
\ee
Finally, we can consider the the energy density $T_{00}(x)$ in an interval $[0,L]$ at a fixed time $t=0$. This decomposes into $T_{++}$ and $T_{--}$ parts as above. For any state where both of these components are uniform on the interval, we have that
\be
T_{00} \ge -{c \pi \over 6 L^2} \; .
\ee
since both $T_{++}$ and $T_{--}$ satisfy (\ref{T0const}). The lower bound is precisely what we have in the vacuum state for a circle of length $L$ with antiperiodic boundary conditions for fermions. \footnote{We have $c=1$ for a Dirac fermion, so the bound on uniform energy density for a segment of width $L$ in a Minkowski space theory (assuming both $T_{++}$ and $T_{--}$ are also uniform) is 
\be
\label{T00const}
T_{00} \ge -{\pi \over 6 L^2} \; .
\ee}

It remains to argue that we can't do any better by considering non-uniform $T_{++}$ and $T_{--}$ that add to give a uniform energy density in the interval $[0,L]$. To show this, suppose that $T_{++}(x^+)$ and $T_{--}(x^-)$ give an energy density with the minimum possible uniform energy $E_0$ on $[-L/2,L/2]$ at $t=0$. Then
\be
T_{++}(s) + T_{--}(-s) = E_0 \qquad \qquad s \in [-L/2,L/2] \; .
\ee
It must be that the quantum mechanics potentials
\be
V_1(s) = \left\{ \ba{ll} T_{++}(s) & \qquad s \in [-L/2,L/2] \cr \infty & \qquad |s| > L/2 \ea \right. \qquad 
V_2(s) = \left\{ \ba{ll} T_{--}(s) & \qquad s \in [-L/2,L/2] \cr \infty & \qquad |s| > L/2 \ea \right.
\ee
each have a ground state with zero energy. Otherwise, the QEI would be violated (if the ground state energy is negative) or (if the energy is positive) we could add a negative constant to $T_{++}$ or $T_{--}$ so that $T_{00}$ is lower and the QEI are still satisfied.

Since the two potentials sum to a constant in $[-L/2,L/2]$ and each have a zero energy ground state, we can write 
\be
T_{++}(s) = V(x) - E_+ \qquad T_{--}(s) = -V(x) - E_-
\ee
where $E_+$ and $E_-$ are the ground state energies for the potentials equal to $\pm V(x)$ in $[-L/2,L/2]$ and infinity outside. Since we are assuming that $E_0 = T_{++}(s) + T_{--}(s) = - E_+ - E_-$ is minimal, the potential $V(x)$ satisfies the conditions of the following lemma and is therefore constant. 
\paragraph{Lemma}
{\it Let $E_{\pm}$ be the ground state energies for potentials
\be
V_\pm(s) = \left\{ \ba{ll} \pm V(s) & \qquad s \in [-L/2,L/2] \cr \infty & \qquad |s| > L/2 \ea \right.
\ee
for some $V(s)$ and suppose that $V(s)$ extremizes $E_+ + E_-$. Then $V(s)$ is constant.}
\paragraph{Proof}
Since $E_+ + E_-$ is extremal, it should vanish under any variation $\delta V(x)$. Using first order quantum perturbation theory, we have that
\be
\delta (E_+ + E_-) = \int dx (|\psi^0_+(x)|^2 - |\psi^0_-(x)|^2 ) \delta V(x) 
\ee
where $\psi^0_{\pm}(x)$ are the ground state wavefunctions for the potentials $V_{\pm}$; these can be taken to be real and positive (the ground state wavefunction has no nodes). Setting this to zero, we have (using that $\psi_{\pm}$ are real and positive) that
\be
\psi^0_+(x) = \psi^0_-(x) \; .
\ee
According to the Schr\"odinger equation (taking $\psi = \psi_+ = \psi_-$), 
\be
-C {d \over ds^2} \psi (s) \pm V(s) \psi(s) = E_{\pm} \psi (s) \; .
\ee
Subtracting the $-$ equation from the $+$ equation, and dividing by $\psi(s)$ in $(-L/2,L/2)$ (where $\psi$ is positive), we get
\be
V(s) = E_+ - E_- \qquad s \in (-L/2,L/2)
\ee
which is constant, as claimed. $\square$

\bibliography{refs}
\bibliographystyle{jhep}

\end{document}